\documentclass[%
 reprint,
 amsmath,amssymb,
 aps,
 pra,
]{revtex4-2}

\usepackage{graphicx}
\usepackage{dcolumn}
\usepackage{bm}
\usepackage{cancel}
\usepackage{xcolor}
\usepackage{hyperref}
\hypersetup{colorlinks=true,allcolors = blue}

\begin{document}

\preprint{APS/123-QED}

\title{Elliptic billiard with harmonic potential: Classical description}

\author{Bernardo Barrera$^1$}
\author{Juan P. Ruz-Cuen$^2$}
\author{Julio C. Guti\'{e}rrez-Vega$^2$}%
\email{juliocesar@tec.mx}
\affiliation{$^1$Department of Physics, Boston University, Boston, Massachusetts 02215, USA}
\affiliation{$^2$Photonics and Mathematical Optics Group, Tecnol\'{o}gico de Monterrey, Monterrey, M\'{e}xico 64849}

\date{\today}

\begin{abstract}
The classical dynamics of the isotropic two-dimensional harmonic oscillator confined by an elliptic hard wall is discussed. The interplay between the harmonic potential with circular symmetry and the boundary with elliptical symmetry does not spoil the separability in elliptic coordinates; however, it generates non-trivial energy and momentum dependencies in the billiard. We analyze the equi-momentum surfaces in the parameters space and classify the kinds of motion the particle can have in the billiard. The winding numbers and periods of the rotational and librational trajectories are analytically calculated and numerically verified. A remarkable finding is the possibility of having degenerate rotational trajectories with the same energy but different second constant of motion and different caustics and periods. The conditions to get these degenerate trajectories are analyzed. Similarly, we show that obtaining two different rotational trajectories with the same period and second constant of motion but different energy is possible.
\end{abstract}

\maketitle


\noindent B. Barrera, J. P. Ruz-Cuen, and J. C. Guti\'errez-Vega, \textit{Elliptic billiard with harmonic potential: Classical description}, Phys. Rev. E \textbf{108}, 034205 (2023).
\href{https://doi.org/10.1103/PhysRevE.108.034205}{https://doi.org/10.1103/PhysRevE.108.034205}

\section{Introduction}

The study of billiards in the classical and quantum regimes is valuable as it provides a simple way to model physical phenomena, for example, particle trapping at the nanometer scale, quantum-classical correspondence, low disorder systems, quantum dots, chaotic systems, laser dynamics in microcavities, ray-optics approximation in waveguides, among others \cite{BirdBOOK,Luna2001,Stone2010,Ponomarenko2008}.

An elliptic billiard consists of a point particle moving inside a planar elliptic domain, bouncing elastically at its hard boundary \cite{TabachnikovBOOK}. Investigation of the mathematical and physical properties of elliptical billiards in both the classical and quantum regimes has a long history \cite{TabachnikovBOOK, GutzwillerBOOK, Berry1981}. It is well-known that the elliptical billiard is an integrable system with two well-defined constants of motion: the energy and the product of angular momenta about the foci \cite{Chang, Waalkens, P05, P16}. The particle moves rectilinearly, forming a polygonal trajectory with vertices on the billiard boundary. The system presents two types of motion: rotational and librational, depending on the sign of the second constant of motion. The trajectories are always tangent to elliptic caustics for rotational motion or hyperbolic caustics for librational motion \cite{P05}.

Various modifications to the elliptical geometry have been studied, such as the transition to oval or circular billiards \cite{Berry1981, Magner}, or the annular \cite{P16} and open boundary structures \cite{P68}. 
Variations of the potential to smooth the sharp elliptical boundary have also been considered \cite{Lynch2019}. 
Most of these properties have been verified experimentally in recent years, thanks to the improvement of nanofabrication processes that allow the construction of quantum corrals to confine electrons \cite{Crommie}. To this day, interesting geometric properties of the trajectories in elliptic billiards continue to be discovered \cite{Garcia, Garcia2}. 

In this paper, we study the dynamic properties of the system formed by the elliptic billiard and the isotropic harmonic potential attracting to the center of the ellipse. The trajectory is, in general, a self-intersecting polygon whose sides are elliptical segments connecting at the boundary. We present a derivation of the second constant of motion and propose a suitable normalization scheme that allows mapping all scenarios in the billiard. The interplay between the harmonic potential with circular symmetry and the boundary with elliptical symmetry does not affect the separability in elliptic coordinates, but it generates non-trivial energy and momentum dependencies in the billiard that are absent in the elliptic billiard without potential. We will therefore discuss the behavior of equi-momentum surfaces in the space of parameters that allow the characterization of the four types of motion the particle can exhibit. We derive the conditions to obtain periodic orbits in the billiard by applying the Hamilton-Jacobi theory \cite{FetterBOOK,GoldsteinBOOK}. The analytical evaluation of the action-angle variables yields closed-form expressions for the winding numbers and the periods of the librational and rotational orbits. From these expressions, several geometric constructions can be developed. A remarkable finding is the possibility of having degenerate rotational trajectories with the same energy but different second constant of motion and caustics and periods. The conditions to get these degenerate trajectories are analyzed. Similarly, we show that obtaining two different rotational trajectories with the same period and second constant of motion but different energy is possible.

From a historical point of view, the antecedents of this problem can be traced back to Jacobi, who, in 1884, analyzed the problem of the motion of a particle along the surface of a triaxial ellipsoid under the action of an elastic force directed toward the center of the ellipsoid \cite{JacobiBOOK}. Suppose one of the axes of the ellipsoid tends to zero. In that case, the Jacobi problem reduces to the problem of the oscillations of the harmonic oscillator inside an ellipse. More recently, Wiersig studied the classical dynamics of the triaxial ellipsoidal billiard with harmonic potential describing the motion in terms of the energy surfaces in the space of action variables \cite{Wiersig}. Dragovi{\'c} et al. extended the study of the ellipsoids to $n$ dimensions but without potential \cite{Dragovic}. 

The most direct antecedent of our work is the analysis by M. Radnovi{\'c}, published in 2015, on elliptic billiards with Hooke's potential \cite{Radnovic2015}.
Radnovi{\'c} uses Fomenko graphs to characterize the billiard's topologies. Her analysis provides expressions for the caustics, their geometric properties, and the bifurcation diagram. In this work,  we apply the Hamilton-Jacobi formalism in elliptic coordinates, which allows for generating many additional analytical results (not reported in Radnovic's paper) such as the Poincar{\'e} maps, the condition for periodic rotational and librational trajectories, the winding number function, the eigen-momentum surfaces, graphs of the trajectories, the analytical expressions for the periods of periodic orbits, among others. Additionally, our analysis reveals that it is possible to have degenerate trajectories with the same energy but different second constants of motion, and we found the condition for this to happen. 

The material in this paper focuses on the classical description of the billiard. It constitutes the first part of a more extended analysis considering the quantum description and the semiclassical approximation. The classical characterization of the elliptic billiard with harmonic potential is sufficiently complex in terms of phenomena and properties that its analysis is justified in separate papers. This work consolidates and extends previous analysis of classical billiards with harmonic potentials \cite{Radnovic2015,KozlovBOOK}.

\section{Statement of the problem}

We consider the motion of a point particle with mass $M$ in an isotropic two-dimensional harmonic potential%
\begin{equation}
U\left(  r\right)  =\frac{1}{2}M\omega^{2}r^{2}=\frac{1}{2}M\omega^{2}\left(
x^{2}+y^{2}\right)  , \label{U}%
\end{equation}
where $\omega$ is the angular frequency of the oscillator. The particle is confined in the region of the plane $(x,y)$ bounded by the ellipse
\begin{equation}
\frac{x^{2}}{a^{2}}+\frac{y^{2}}{b^{2}}=1,\qquad\qquad\left(  b\leq a\right),
\end{equation}
whose foci are located at $x_{\pm}=\pm f=\pm(a^{2}-b^{2})^{1/2}$, as shown in Fig. 1.%

\begin{figure}[t]
\includegraphics[width=8cm]{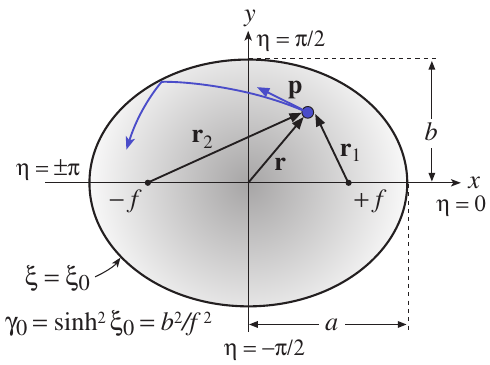}
\caption{\label{Fig01} Geometry of the elliptic billiard with harmonic potential. For a
given focal distance $f,$ the boundary is defined by the \textit{radial}
elliptic coordinate $\xi=\xi_{0}$ or, alternatively, by the parameter
$\gamma_{0}=\left(  b/f\right)  ^{2}=\sinh^{2}\xi_{0}.$ }
\end{figure}

The particle moves inside the billiard under the effect of the central force produced by the parabolic potential. As it travels through the potential, the total (kinetic plus potential) energy
\begin{equation}
E=\frac{p^{2}}{2M}+\frac{1}{2}M\omega^{2}r^{2}=\mathrm{cons}\geq0, \label{E}%
\end{equation}
and the angular momentum about the origin
\begin{equation}
\mathbf{L}=\mathbf{r}\times\mathbf{p}=L\widehat{\mathbf{z}}=\left(
xp_{y}-yp_{x}\right)  \widehat{\mathbf{z}}, \label{L}%
\end{equation}
remain constant along the trajectory. Here, $\mathbf{p}$ is the momentum of the particle, and $p,~p_{x},~p_{y}$ are their magnitude and Cartesian components, respectively.

If the particle does not hit the boundary, it is well-known that its orbit is a closed ellipse centered at the origin whose size and orientation are determined by the initial conditions \cite{FetterBOOK,GoldsteinBOOK}.

On the other hand, if the particle hits the boundary, it makes a polygonal trajectory with elliptical segments connecting at the reflection points. In this case, the energy $E$ before and after each impact is still conserved because the collisions are elastic, but the angular momentum $L$ changes because the force exerted by the elliptic wall on the particle is not central. From the analysis of the elliptic billiard with zero potential \cite{Chang,Waalkens,P05}, it is known that the quantity that is conserved in the reflection with the elliptic wall is the product of the angular momenta about the foci, i.e.,
\begin{equation}
\mathbf{L}_{1}\cdot\mathbf{L}_{2}=\left(  \mathbf{r}_{1}\times\mathbf{p}%
\right)  \cdot\left(  \mathbf{r}_{2}\times\mathbf{p}\right)  =L_{1}L_{2},
\label{L1pL2}%
\end{equation}
where $\mathbf{r}_{1}=\left(  x-f\right)  \widehat{\mathbf{x}}+y\widehat
{\mathbf{y}}$ and $\mathbf{r}_{2}=\left(  x+f\right)  \widehat{\mathbf{x}%
}+y\widehat{\mathbf{y}},$ see Fig. 1. By expanding Eq. (\ref{L1pL2}), the
product $L_{1}L_{2}$ can be easily related to the angular momentum $L$ as
follows:
\begin{equation}
L_{1}L_{2}=L^{2}-f^{2}p_{y}^{2}, \label{L1L2}%
\end{equation}
where $p_{y}$ is the component of the momentum along the $y$ axis.

If the potential energy $U(r)$ were zero in all points of the billiard's area, the quantity $L_{1}L_{2}$ would be conserved as the particle moves rectilinearly inside the billiard. However, as the particle moves elliptically within the parabolic potential, $L_{1}L_{2}$ is not constant anymore; thus, it cannot serve as a second constant of motion for our problem. It is then necessary to identify the second constant of motion needed to characterize the elliptic billiard with parabolic potential. 

\section{Derivation of the second constant of motion}

We begin by noting the total energy of the particle [Eq. (\ref{E})] can be split into two Cartesian contributions
\begin{align}
E  &  =E_{x}+E_{y}\nonumber\\
&  =\left(  \frac{p_{x}^{2}}{2M}+\frac{M\omega^{2}x^{2}}{2}\right)  +\left(
\frac{p_{y}^{2}}{2M}+\frac{M\omega^{2}y^{2}}{2}\right)  .
\end{align}
Both $E_{x}$ and $E_{y}$ are individually conserved during the motion through the harmonic potential. Now, when the particle hits the boundary, the values of $E_{x}$ and $E_{y}$ of the incident trajectory shift an amount $W,$ i.e.
\begin{equation}
E_{x}\rightarrow E_{x}-W,\qquad\qquad E_{y}\rightarrow E_{y}+W, \label{ExEy}%
\end{equation}
such the total energy $E=E_{x}+E_{y}$ remains constant after the collision. The shift $W$ is attributable to a rearrangement of the kinetic energy contributions among the $x$ and $y$ components since the potential energy is the same before and after the impact.

To find $W,$ we recall that $L_{1}L_{2}$ does not change at the reflection with the boundary, i.e., $\Delta\left(  L_{1}L_{2}\right)  =0$, then from Eq.
(\ref{L1L2}) we have%
\begin{equation}
\Delta\left(  L^{2}\right)  = \cancelto{0}{\Delta\left(  L_{1}L_{2}\right) }
 + f^{2} \Delta\left(  p_{y}^{2}\right) =  2Mf^{2}W,
\end{equation}
where we applied $\Delta\left(  p_{y}^{2}\right)  =2M\left(  \Delta
E_{y}\right)  =2MW.$ Replacing $W$ in Eqs. (\ref{ExEy}) gives%
\begin{equation}
E_{x}\rightarrow E_{x}-\frac{\Delta\left(  L^{2}\right)  }{2Mf^{2}},\qquad
E_{y}\rightarrow E_{y}+\frac{\Delta\left(  L^{2}\right)  }{2Mf^{2}}.
\end{equation}

To construct two conserved quantities, we compensate $E_{x}$ and $E_{y}$ by the amount of energy that is lost (gained)\ at the collision, that is%
\begin{equation}
\mathcal{E}_{x}\equiv E_{x}+\frac{L^{2}}{2Mf^{2}},\qquad\mathcal{E}_{y}\equiv
E_{y}-\frac{L^{2}}{2Mf^{2}}.
\end{equation}
Both $\mathcal{E}_{x}$ and $\mathcal{E}_{y}$ remain constant at (a) each collision with the elliptic boundary and (b) along the segments between collisions because $E_{x}$, $E_{y},$ and $L$ are conserved quantities in the harmonic potential.

We can set combinations of $\mathcal{E}_{x}$ and $\mathcal{E}_{y}$ that are also conserved quantities themselves. For instance
\begin{align}
\mathcal{E}_{x}+\mathcal{E}_{y}  &  =E,\\
\mathcal{E}_{x}-\mathcal{E}_{y}  &  =E_{x}-E_{y}+\frac{L^{2}}{Mf^{2}}.
\label{Lmen}%
\end{align}
The first quantity is evidently the total energy of the particle. The second
quantity Eq. (\ref{Lmen}) can be rewritten, using Eq. (\ref{L1L2}) and multiplying
by $Mf^{2}$, in the following form%
\begin{equation}
\Gamma\equiv L_{1}L_{2}-f^{2}M^{2}\omega^{2}y^{2}=\mathrm{cons,}
\label{Lambda}%
\end{equation}
where $\Gamma$ has units of squared angular momentum.

Throughout the paper, we will consider the total energy $E$ [Eq. (\ref{E})] and the quantity $\Gamma$ as the two fundamental constants of motion of the billiard. We choose the form of Eq. (\ref{Lambda}) because the parameters $f$ and $\omega$ appear explicitly as simple factors, allowing us to easily make the transition to the elliptic billiard without potential (if $\omega \rightarrow0$ then $\Gamma\rightarrow L_{1}L_{2}$), or to the case of the circular billiard with harmonic potential (if $f\rightarrow0$ then $\Gamma\rightarrow L^{2}$). 

\section{Formulation in elliptic coordinates}

The problem is conveniently described in elliptic coordinates%
\begin{equation}
x=f\cosh\xi\cos\eta,\text{\qquad}y=f\sinh\xi\sin\eta,
\end{equation}
where $\xi\in\left[  0,\infty\right)  $ is the \textit{elliptic radial} coordinate and $\eta\in\left(  -\pi,\pi\right]  $ is the \textit{elliptic angular} coordinate. Lines of constant $\xi$ are confocal ellipses and lines of constant $\eta$ are confocal hyperbolae. The locus $\xi=0$ corresponds to the interfocal line $\left\vert x\right\vert \leq f.$ 

The surface of the billiard is specified by the region
\begin{equation}
\xi\in\left[  0,\xi_{0}\right]  ,\qquad\qquad\eta\in\left(  -\pi,\pi\right]  ,
\end{equation}
where
\begin{equation}
\xi=\xi_{0}=\mathrm{arctanh}\left(  b/a\right)  ,
\end{equation}
defines the elliptic boundary.

The constants of motion $E$ and $\Gamma$ can be expressed in terms of the
elliptical coordinates $\left(  \xi,\eta\right)  $ and canonical momenta
$\left(  p_{\xi},p_{\eta}\right)  ,$ where $p_{\xi}$ and $p_{\eta}$ are the
\textit{radial} and \textit{angular} components of the momentum vector in
elliptic coordinates, i.e.,
\begin{equation}
\mathbf{p}=M\mathbf{v}=\left(  \frac{p_{\xi}}{\sigma}\right)  \mathbf{\hat
{\xi}}+\left(  \frac{p_{\eta}}{\sigma}\right)  \mathbf{\hat{\eta}}, \label{pV}%
\end{equation}
with%
\begin{equation}
\sigma=\sigma\left(  \xi,\eta\right)  =f\sqrt{\cosh^{2}\xi-\cos^{2}\eta},
\end{equation}
being the scaling factor of the elliptic coordinates. The canonical momenta $p_\xi$ and $p_\eta$
have units of momentum per length, that is, angular momentum.
In elliptic coordinates, the total energy $E$ [Eq. (\ref{E})] becomes%
\begin{equation}
E=\frac{p_{\xi}^{2}+p_{\eta}^{2}}{2M\sigma^{2}}+\frac{M\omega^{2}f^{2}}%
{2}\left(  \cosh^{2}\xi-\sin^{2}\eta\right)  , \label{EE}%
\end{equation}
where we applied $r^{2}=x^{2}+y^{2}=f^{2}(\cosh^{2}\xi-\sin^{2}\eta),$ and the
constant $\Gamma$ [Eq. (\ref{Lambda})] becomes%
\begin{equation}
\Gamma=\frac{f^{2}}{\sigma^{2}}\left(  p_{\eta}^{2}\sinh^{2}{\xi}-p_{\xi}%
^{2}\sin^{2}{\eta}\right)  -f^{4}M^{2}\omega^{2}\sinh^{2}{\xi}\sin^{2}{\eta.}
\label{G}%
\end{equation}

Inspection of Eq. (\ref{Lambda}) or (\ref{G}) reveals that $\Gamma$ minimizes when the particle moves along the $y$ axis. By replacing $\eta=\pm\pi/2$ and $p_{\eta}=0,$ we get after some calculations $\Gamma_{\min}=-2MEf^{2},$ where $E$ is the total energy. On the other hand, the maximum value of $\Gamma$ occurs when the particle moves tangentially along the elliptic boundary. In this case, $\xi=\xi_{0}$ and $p_{\xi}=0,$ and we obtain $\Gamma_{\max }=2MEb^{2}.$ Consequently, the range of $\Gamma$ is given by 
\begin{equation}
\Gamma\in\left[  -2Mf^{2}E,2Mb^{2}E\right]  .
\end{equation}

At this point, it is convenient to introduce a normalized version of $\Gamma$ that will serve as a new dimensionless constant of motion, namely
\begin{equation}
\gamma=\frac{\Gamma}{\Gamma_0}\equiv\frac{\Gamma}{2Mf^{2}E}\in\left[-1,\gamma_{0}\right]  ,\label{g}%
\end{equation}
where%
\begin{equation}
\Gamma_0\equiv2Mf^{2}E,\qquad\gamma_{0}\equiv\frac{b^{2}}{f^{2}}=\sinh^{2}\xi
_{0},
\end{equation}
Note that the upper limit $\gamma_{0}$ is defined only by the geometric parameters of the billiard boundary. In fact, similar to the eccentricity $\epsilon=f/a=\mathrm{sech}\xi_{0}$, the parameter $\gamma_{0}$ could be used to specify the ellipticity of the boundary. 

We now combine Eqs. (\ref{EE}) and (\ref{G}) to decouple the momenta $p_{\xi}$ and 
$p_{\eta}$. After some algebraic manipulations, we get%
\begin{align}
p_{\xi}^{2}  &  =\Gamma_0\left(  \sinh^{2}\xi-\gamma-\beta\sinh^{2}\xi\cosh
^{2}\xi\right)  ,\label{pe}\\
p_{\eta}^{2}  &  =\Gamma_0\left(  \sin^{2}\eta+\gamma-\beta\sin^{2}\eta\cos
^{2}\eta\right)  , \label{pn}%
\end{align}
where 
\begin{equation}
\beta\equiv\frac{M\omega^{2}f^{2}}{2E}=\frac{U_{f}}{E},\qquad\qquad
U_{f}=\frac{1}{2}M\omega^{2}f^{2},
\end{equation}
is a new dimensionless constant of motion associated with the energy, and
$U_f$ is the potential energy at a radius equal to $f$. 

Each of the equations (\ref{pe}) and (\ref{pn}) can be interpreted as a Hamiltonian system with one degree of freedom, with effective potential
$U_{\mathrm{eff}}\left(  \xi\right)  \propto \beta\sinh^{2}\xi\cosh^{2}\xi -\sinh^{2}\xi$, 
and $U_{\mathrm{eff}}\left(  \eta\right)  \propto \beta\sin^{2}\eta\cos^{2}\eta -\sin^{2}\eta$, respectively.

The constant of motion $\beta$ characterizes how strong the coupling of the particle to the harmonic potential is.  It accounts for the effects of the energy $E$ (a dynamic parameter), the mass $M$ (particle's property), the frequency $\omega$ (potential's property), and the distance $f$ (elliptic boundary's property).  

Low values of $\beta$ correspond to a small coupling. In this case the particle moves inside the billiard almost as if it were a free particle, and thus their trajectory segments become quasi-straight lines that collide with the boundary. If $\beta=0$, the system reduces to the well-known elliptic billiard with a free particle inside \cite{Chang,Waalkens,P05}. 

On the other hand, high values of $\beta$ correspond to a strong coupling where the particle's excursion around the origin is small. In this case the trajectory does not reach the boundary and thus becomes a closed ellipse centered at the origin \cite{FetterBOOK,GoldsteinBOOK}.   

Alternatively, $\beta$ could be expressed as $\beta=f^{2}/R^{2},$ where $R=\sqrt{2E/M\omega^{2}}$ is the amplitude that a one-dimensional harmonic oscillator with energy $E$ reaches in the parabolic potential. In other words, $R$ defines the largest circular region where the particle could move for a given energy $E$ if there were no elliptical wall.  

In what follows, the parameters $\gamma$ and $\beta$ will be considered as the constants of motion of the problem. $\beta$ is related to the energy, and $\gamma$ to the constant $\Gamma.$%

\begin{figure}[t]
\includegraphics[width=8cm]{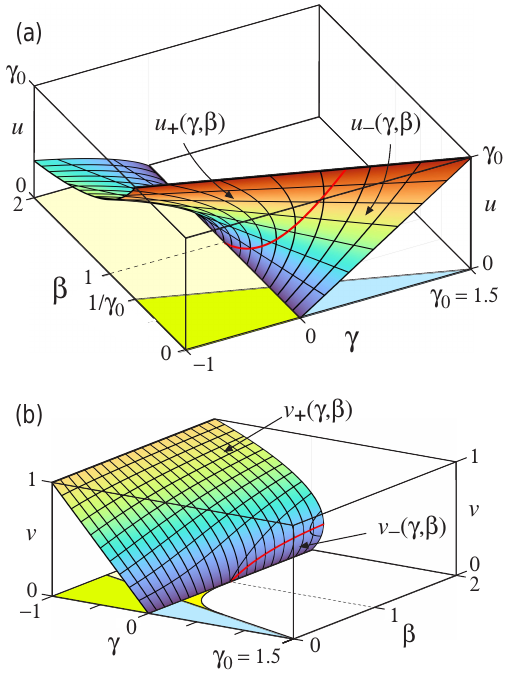}
\caption{\label{Fig2} a) Surfaces $u_{\pm}\left(  \gamma,\beta\right)  $ for $\gamma
\in\left[  -1,\gamma_{0}\right]  ,$ $\beta\in\left[  0,2\right]  ,$ and
$u\in\left[  0,\gamma_{0}\right]  $ with $\gamma_{0}=1.5.$ Red curve is the
branch line Eq. (\ref{curveu}). The surface is doubled-valued in the region
defined by Eq. (\ref{E0}). (b) Surfaces $v_{\pm}\left(  \gamma,\beta\right)  $
in the interval $v\in\left[  0,1\right]  $. Red curve is the branch line Eq.
(\ref{curvev}). The surface is doubled-valued in the region defined by Eqs.
(\ref{vd}).}
\end{figure}

\section{Equi-momentum surfaces and classification of the trajectories}

Equations (\ref{pe}) and (\ref{pn}) describe the dynamics of the particle in the billiard. To facilitate their analysis we rewrite them in the form
\begin{align}
p_{\xi}^{2}\left(  \gamma,\beta;u\right)   &  =\Gamma_0\left[  -\beta
u^{2}+\left(  1-\beta\right)  u-\gamma\right]  ,\label{pee}\\
p_{\eta}^{2}\left(  \gamma,\beta;v\right)   &  =\Gamma_0\left[  \beta
v^{2}+\left(  1-\beta\right)  v+\gamma\right]  , \label{pnn}%
\end{align}
where
\begin{equation}
u\equiv\sinh^{2}\xi,\qquad\text{and}\qquad v\equiv\sin^{2}\eta.
\end{equation}

The domains of the square canonical momenta $p_{\xi}^{2}$ and $p_{\eta}^{2}$ in the three-dimensional spaces $\left(  \gamma,\beta;u\right)  $ and $\left(
\gamma,\beta;v\right)  $ are
\begin{equation}%
\begin{array}
[c]{lll}%
\gamma\in\left[  -1,\gamma_{0}\right]  , &  & \beta\in\lbrack0,\infty),\\
u\in\left[  0,\gamma_{0}\right]  , &  & v\in\left[  0,1\right]  ,
\end{array}
\label{rango}%
\end{equation}
where $\gamma_{0}=b^{2}/f^{2}=\sinh^{2}\xi_{0}$ [Eq. (\ref{g})]. Note that the normalization scheme makes the upper limits of $\gamma$ and $u$ equal to $\gamma_{0}$.

To ensure that $p_{\xi}$ and $p_{\eta}$ are real quantities, the triplets
$\left(  \gamma,\beta;u\right)  $ and $\left(  \gamma,\beta;v\right)  $ must
lie within the regions%
\begin{align}
p_{\xi}^{2} &  \geq0\quad\rightarrow\quad-\beta u^{2}+\left(  1-\beta\right)u-\gamma\geq0,\label{pe2}\\
p_{\eta}^{2} &  \geq0\quad\rightarrow\quad\quad\beta v^{2}+\left(1-\beta\right)  v+\gamma\geq0.\label{pn2}%
\end{align}
We now analyze each condition separately.

\subsection{Surface $p_{\xi}^{2}\left(  \gamma,\beta;u\right)  =0.$}

Let us first discuss the \textit{radial} condition Eq. (\ref{pe2}). The locus
$p_{\xi}^{2}\left(  \gamma,\beta;u\right)  =0$ corresponds to an
\textit{equi-momentum} surface in the three-dimensional parametric space $\left(
\gamma,\beta;u\right)  $ where all points on the surface have zero
\textit{radial} momentum. This surface separates the valid region $p_{\xi}%
^{2}>0$ from the forbidden region $p_{\xi}^{2}<0$. Solving the quadratic
equation for $u$, we see that the surface $p_{\xi}^{2}\left(  \gamma
,\beta;u\right)  =0$ is composed of two sheets given by%
\begin{equation}
u_{\pm}\left(  \gamma,\beta\right)  =\frac{1-\beta\pm\mathcal{D}}{2\beta},
\label{umm}%
\end{equation}
where
\begin{equation}
\mathcal{D}=\mathcal{D}\left(  \gamma,\beta\right)  =\sqrt{\left(
\beta-1\right)  ^{2}-4\gamma\beta}. \label{D}%
\end{equation}

Since $\mathcal{D}$ has to be real, then $\gamma$ and $\beta$ must satisfy the
condition
\begin{equation}
\gamma\leq\frac{\left(\beta-1\right)^{2}}{4\beta}. \label{gb}%
\end{equation}

The behaviors of $u_{\pm}\left(  \gamma,\beta\right)$ are shown in Fig. 2(a)
for the valid ranges of the variables $\left(  \gamma,\beta;u\right)  $ in Eq.
(\ref{rango}). The surfaces $u_{+}\left(  \gamma,\beta\right)  $ and
$u_{-}\left(  \gamma,\beta\right)  $ bifurcate at the curve%
\begin{equation}
\left(  \gamma,\beta,u\right)  =\left(  \frac{\left(  \beta-1\right)  ^{2}}{4\beta},\beta,\frac{1-\beta}{2\beta}\right)  ,\quad\beta\geq0,
\label{curveu}%
\end{equation}
see the red line in Fig. 2(a).

Now, the variable $u$ is limited to the range $\left[  0,\gamma_{0}\right]  .$
The intersection of the plane $u=\gamma_{0}$ with the surfaces $u_{\pm}\left(
\gamma,\beta\right)  $ occurs at straight line%
\begin{equation}
\beta=\frac{\gamma_{0}-\gamma}{\gamma_{0}\left(  1+\gamma_{0}\right)  },
\label{sl}%
\end{equation}
that goes from the point $\left(  -1,1/\gamma_{0};\gamma_{0}\right)  $ to the
point $\left(  \gamma_{0},0;\gamma_{0}\right)  .$ The intersection of the
plane $u=0$ with the surfaces $u_{\pm}\left(  \gamma,\beta\right)  $
corresponds to the line $\gamma=0$, i.e., the $\beta$ axis in the space
$\left(  \gamma,\beta;u\right)  .$

\subsection{Surface $p_{\eta}^{2}\left(  \gamma,\beta;v\right)  =0.$}

The equi-momentum surfaces $p_{\eta}^{2}\left(  \gamma,\beta;v\right)  =0$ are
obtained by solving Eq. (\ref{pn2}), we get%
\begin{equation}
v_{\pm}\left(  \gamma,\beta\right)  =\frac{\beta-1\pm\mathcal{D}}{2\beta},
\label{vmm}%
\end{equation}
where $\mathcal{D}$ is given by Eq. (\ref{D}).

Since the argument of the radical in $\mathcal{D}$ is the same as Eq.
(\ref{umm}), the condition in Eq. (\ref{gb}) applies for this case as well.
The curve where the surfaces $v_{+}\left(  \gamma,\beta\right)  $ and
$v_{-}\left(  \gamma,\beta\right)  $ bifurcate is%
\begin{equation}
\left(  \gamma,\beta,v\right)  =\left(  \frac{\left(  \beta-1\right)  ^{2}%
}{4\beta},\beta,\frac{\beta-1}{2\beta}\right)  ,\quad\beta\geq0,
\label{curvev}%
\end{equation}
which is the reflection of the curve (\ref{curveu}) on the plane $u=0,$ as
shown in Fig. 2(b). This result comes from the fact that
\begin{equation}
v_{\pm}\left(  \gamma,\beta\right)  =-u_{\mp}\left(  \gamma,\beta\right)  ,
\end{equation}
as can be corroborated from Eqs. (\ref{umm}) and (\ref{vmm}).

The equi-momentum surface $v(\beta,\gamma)$ is double-valued at the region defined by
\begin{equation}
\gamma\in\left[  0,\frac{\left(  \beta-1\right)  ^{2}}{4\beta}\right]
,\qquad\beta>1. \label{vd}%
\end{equation}

The range of $v$ is $\left[  0,1\right]  .$ The intersection of the planes
$v=0$ and $v=1$ with the surfaces $v_{\pm}\left(  \gamma,\beta\right)  $ are
the straight lines $\gamma=0$ and $\gamma=-1,$ respectively.%

\begin{figure}[t]
\includegraphics[width=8cm]{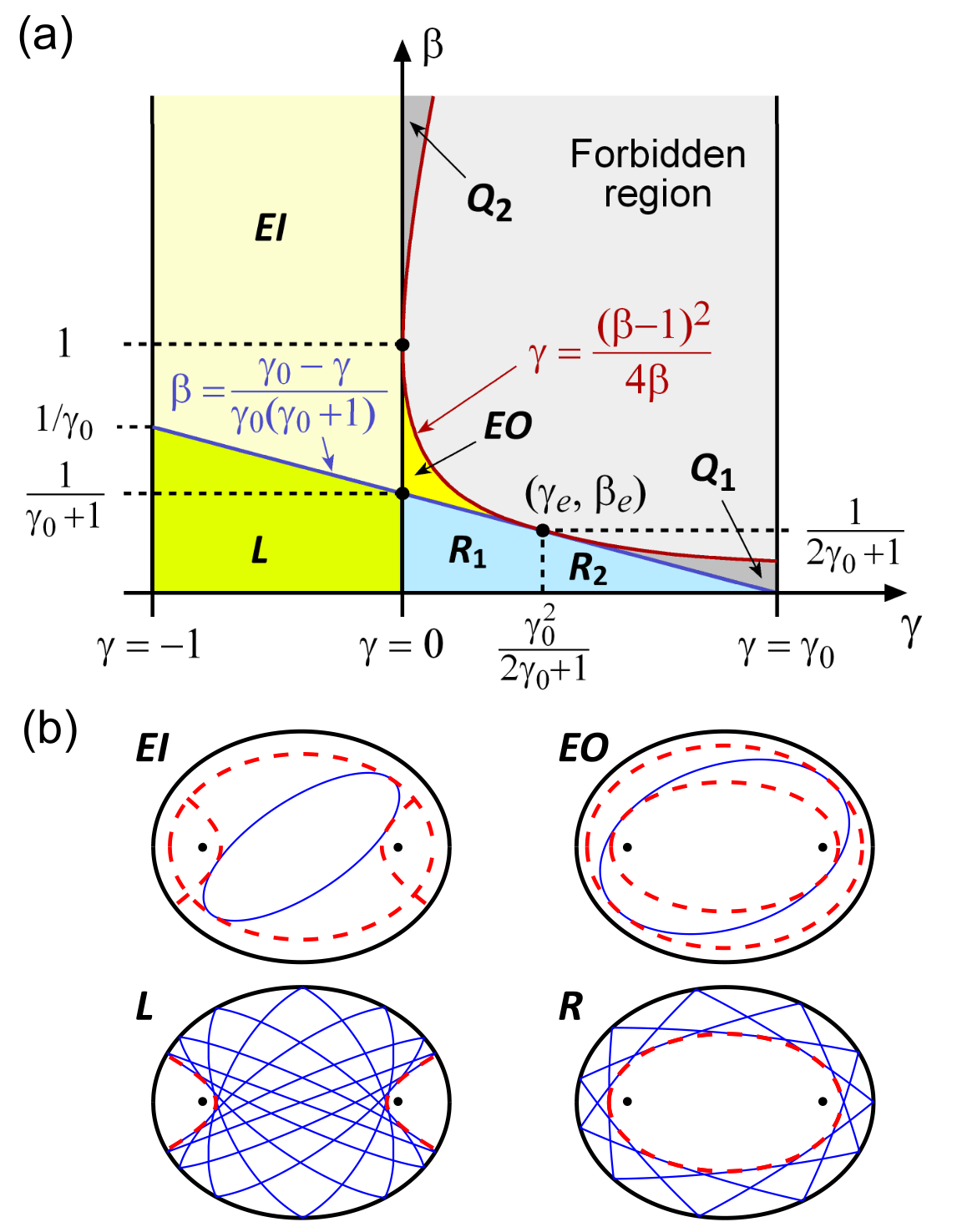}
\caption{\label{Fig3} (a) Regions on the plane $\left(  \gamma,\beta\right)  $
corresponding to different types of trajectories in the billiard with
$\gamma_{0}=1.5.$ (b) Representative periodic trajectories and their caustics
(red dashed lines). Orbits $\left(  3,7\right)  $ for $\mathcal{L}$- and
$\mathcal{R}$-motions.}
\end{figure}

\subsection{Classification of the trajectories \label{Sect_Traj}}

The results discussed above allow us to classify the kinds of orbits the
particle can exhibit in the billiard. As shown in Fig. 3, the plane $\left(
\gamma,\beta\right)  $ is divided into zones by three curves:

\begin{itemize}
\item The (red) curve $\gamma=\left(  \beta-1\right)  ^{2}/4\beta$ separates
the \textit{valid region} of pairs $\left(\gamma,\beta\right)$ that generate allowed trajectories within the billiard, from the \textit{forbidden region} where $p_{\xi}$ and $p_{\eta}$ become imaginary.
The curve is tangent to the $\beta$ axis at the point $\left(0,1\right) .$

\item The (blue) straight line Eq. (\ref{sl}) separates the regions where the
particle hits the boundary (regions below the line) from the regions where the
particle does not hit the boundary (regions above the line). The straight
line is tangent to the curve at the point
\begin{equation}
\left(  \gamma_{e},\beta_{e}\right)  =\left(  \frac{\gamma_{0}^{2}}%
{2\gamma_{0}+1},\frac{1}{2\gamma_{0}+1}\right)  , \label{gebe}%
\end{equation}
as shown in Fig. 3(a).

\item The vertical axis $\gamma=0$ separates the regions where the particle
always crosses the $x$ axis between the foci of the elliptic boundary
(negative $\gamma$) from the regions where it crosses the $x$ axis outside the
interfocal line (positive $\gamma$). If $\gamma=0,$ the particle successively
passes through the foci of the elliptic boundary.
\end{itemize}

\medskip Each zone in the plane $\left(  \gamma,\beta\right)  $ corresponds to
a kind of orbit. \medskip

\textbf{Rotational (}$\mathcal{R}=\mathcal{R}_{1}+\mathcal{R}_{2}$\textbf{)}.
Triangular region defined by%
\begin{equation}
0<\gamma\leq\gamma_{0},\qquad0\leq\beta\leq\frac{\gamma_{0}-\gamma}{\gamma
_{0}\left(  1+\gamma_{0}\right)  },
\end{equation}
as shown in Fig. 3(a). For a point $\left(  \gamma,\beta\right)  $ lying in the
$\mathcal{R}$ region, there is one real root for $p_{\xi}^{2}\left(  u\right)
$ in the interval $u\in\left[  0,\gamma_{0}\right]  $ and none for $p_{\eta
}^{2}\left(  \eta\right)  .$ The particle rotates around the interfocal line
making a polygonal trajectory with vertices at the collision points on the
boundary; see Fig. 3(b). The particle always crosses the $x$ axis outside the
interfocal line. All segments of the trajectory are tangent to a confocal
elliptic caustic $\xi=\xi_{C}$. These points of tangency with the caustic are
just where the \textit{radial} momentum vanishes, i.e., $p_{\xi}^{2}\left(
\xi_{C}\right)  =0.$ From Eq. (\ref{umm}) we obtain
\begin{equation}
\xi_{C}=\mathrm{arcsinh}\left(  \sqrt{u_{-}}\right)  =\mathrm{arcsinh}\left(
\sqrt{\frac{1-\beta-\mathcal{D}}{2\beta}}\right)  , \label{uCL}%
\end{equation}
where $\mathcal{D}$ is given by Eq. (\ref{D}). The \textit{radial} coordinate
of the particle is restricted to the range $\xi\in\left[  \xi_{C},\xi
_{0}\right]  ,$ whereas the \textit{angular} coordinate $\eta$ is not bounded.
As shown in Fig. 3, the region $\mathcal{R}$ is divided into two subregions,
$\mathcal{R}_{1}$ and $\mathcal{R}_{2}$, by the vertical line $\gamma
=\gamma_{e}$. The difference between both regions will be discussed
later.\medskip

\textbf{Librational (}$\mathcal{L}$\textbf{).} Trapezoidal region defined by
\begin{equation}
-1\leq\gamma<0,\qquad0\leq\beta\leq\frac{\gamma_{0}-\gamma}{\gamma_{0}\left(
1+\gamma_{0}\right)  }.
\end{equation}
For a point $\left(  \gamma,\beta\right)  $ lying in this region, there is one
real root for $p_{\eta}^{2}\left(  v\right)  $ in the interval $v\in\left[
0,1\right]  ,$ and none for $p_{\xi}^{2}\left(  \xi\right)  $. The particle
bounces alternately between the top and bottom of the boundary, crossing the
$x$ axis always between the two foci, see Fig. 3(b). Recalling that
$v=\sin^{2}\eta,$ the \textit{angular} momentum $p_{\eta}$ vanishes at
$\eta=\pm\eta_{C}$ and $\eta=\pm\left(  \pi-\eta_{C}\right)  ,$ where
$\eta_{C}\in(0,\pi/2).$ Thus, the librational orbits are confined within two
confocal hyperbolic caustics, as shown in Fig. 3(b). The value of $\eta_{C}$
is obtained with Eq. (\ref{vmm}), namely%
\begin{equation}
\eta_{C}=\arcsin\left(  \sqrt{v_{+}}\right)  =\arcsin\left(  \sqrt{\frac
{\beta-1+\mathcal{D}}{2\beta}}\right)  , \label{vcL}%
\end{equation}
The range of the \textit{radial} coordinate in the librational motion is full,
i.e., $\xi\in\left[  0,\xi_{0}\right]  $.\medskip

\textbf{Elliptical Inner (}$\mathcal{EI}$\textbf{).} Region defined by%
\begin{equation}
-1\leq\gamma<0,\qquad\frac{\gamma_{0}-\gamma}{\gamma_{0}\left(  1+\gamma
_{0}\right)  }<\beta<\infty.
\end{equation}
For a point $\left(  \gamma,\beta\right)  $ lying in this region, there is one
root for $p_{\xi}^{2}\left(  u\right)  $ and one root for $p_{\eta}^{2}\left(
v\right)  .$ The particle trajectory is an ellipse that does not touch the
billiard boundary. Both foci of the billiard are located outside the particle
orbit; thus, it always crosses the $x$-axis within the interfocal line. The trajectory is confined by two hyperbolic caustics and one elliptic caustic.
The hyperbolic caustics are the same as in the librational case, i.e., Eq.
(\ref{vcL}). The elliptic caustic is determined with Eq. (\ref{umm}) taking the positive sign, namely%
\begin{equation}
\xi_{C}^{\mathcal{EI}}=\mathrm{arcsinh}\left(  \sqrt{u_{+}}\right)
=\mathrm{arcsinh}\left(  \sqrt{\frac{1-\beta+\mathcal{D}}{2\beta}}\right)  .
\label{uCEI}%
\end{equation}
\medskip

\textbf{Elliptical Outer (}$\mathcal{EO}$\textbf{).} Region enclosed by the
three lines%
\begin{equation}
\gamma=0,\quad\beta=\frac{\gamma_{0}-\gamma}{\gamma_{0}\left(  1+\gamma
_{0}\right)  },\quad\gamma=\frac{\left(  \beta-1\right)  ^{2}}{4\beta}.
\label{E0}%
\end{equation}
For a point $\left(  \gamma,\beta\right)  $ lying in this region, there are
two different real roots for $p_{\xi}^{2}\left(  u\right)  $ in the interval
$\left[  0,1\right]  ,$ and none for $p_{\eta}\left(  \eta\right)  .$ The
trajectory is again an ellipse that does not touch the billiard boundary, but
now the foci of the billiard are located inside the particle's orbit; thus, it
always crosses the $x$ axis outside the interfocal line. The trajectory is
confined by two elliptical confocal caustics which can be calculated with Eqs.
(\ref{uCL}) and (\ref{uCEI}). \medskip

\textbf{Vertical (}$\mathcal{V}$\textbf{).} If $\gamma=-1$, the particle
becomes a one-dimensional harmonic oscillator moving vertically along the $y$
axis. If $\beta\leq1/\gamma_{0}$, the particle bounces at the covertex points
of the elliptic boundary located at $y=\pm b.$ If $\beta>1/\gamma_{0}$ the
oscillator does not touch the boundary.\medskip

\textbf{Focal (}$\mathcal{F}$\textbf{).} The line $\gamma=0$ is the separatrix
between the librational and the rotational motions. In this case, the particle
crosses through one of the foci, then bounces off the boundary and crosses
through the other focal point, and continues like that, crossing both foci
alternately. As the particle bounces back and forth, the trajectories become
more and more horizontal, and the orbit tends to align with the $x$ axis.
Eventually, the orbit is practically a horizontal harmonic oscillator after
many bounces. When the particle moves along the $x$ axis, if $\beta
\leq1/(1+\gamma_{0})$ it bounces at the vertex points of the boundary located
at $x=\pm a;$ else if $\beta>1/(1+\gamma_{0}),$ the horizontal oscillator does
not reach the boundary.

\medskip

Special points of the plane $\left(  \gamma.\beta\right)  $:\medskip

\begin{itemize}
\item The point $\left(  \gamma_{e},\beta_{e}\right)  $ [Eq. (\ref{gebe})] is
the limiting case when the two elliptic caustics of the $\mathcal{EO}$ motion
collapse into a single caustic equal to the boundary. In this case, the
particle moves tangentially to the boundary \textit{without touching it}. In
other words, if we remove the wall, the particle would continue moving on an
ellipse identical to the billiard boundary due \textit{exclusively} to the attractive
force of the harmonic potential.

\item The point $(0,1)$ is the limiting case when the two elliptic caustics of
the $\mathcal{EO}$ motion collapse into the interfocal line. Then, the
particle becomes a one-dimensional harmonic oscillator moving horizontally
with amplitude $f.$

\item The point $\left(  0,1/(\gamma_{0}+1)\right)  $ is the meeting point of
the four regions $\mathcal{L}$, $\mathcal{R}$, $\mathcal{EI}$, $\mathcal{EO}$,
and can be considered the borderline case of the four types of motion. In this
case, the particle oscillates harmonically along the ellipse's major axis
with amplitude $a$; that is, it \textit{only} touches the elliptical boundary
at their vertex points. Any slight perturbation of this condition leads the
particle to have one of the four main types of motion.

\item A point $\left(  \gamma,\beta\right)  $ lying in the region $Q_{1}$ (see
Fig. 3) produces real positives values of $u_{\pm}$ but both are outside of
the valid interval $u\in\left[  0,\gamma_{0}\right]  $. Then, there are not
possible trajectories in this region.

\item A point $\left(  \gamma,\beta\right)  $ lying in the region $Q_{2}$
leads to negative values of $p_{\xi}^{2},$ so there are no physically valid
solutions in that region either.
\end{itemize}

\subsection{Poincar\'{e} maps}

Figure 4 shows the Poincar\'{e} phase maps in the radial $\left(  \xi,p_{\xi
}\right)  $ and angular $\left(  \eta,p_{\eta}\right)$ position-momentum
spaces for several values of $\beta$. The level curves
correspond to constant values of $\gamma$ in Eqs. (\ref{pe}) and (\ref{pn}). 
These expressions are doubled-valued functions corresponding to the two
possible signs of the momenta. The particle moves in the phase map in a
trajectory where $\gamma$ and $\beta$ (i.e., the energy $E$ and the quantity
$\Gamma$) remain constant. 

We chose the values of $\beta$ to illustrate the typical phase map for each region in the trajectory chart in Fig. 3.
If $\beta=0,$ we recover the known phase maps of the elliptic billiard without potential \cite{P05,P16}. 
The thick black line corresponds to the separatrix $\gamma=0$. 
Note in the maps $\gamma(\xi,p_\xi)$ that the area with positive $\gamma$ decreases as $\beta$ increases. When $\gamma=1/(\gamma_0+1)=0.4$, the thick black line no longer touches the boundary $\xi=\xi_0$, which means that rotational trajectories can no longer exist in the billiard. In the interval $\beta \in [0.4,1]$, the region with positive $\gamma$ corresponds to the $\mathcal{EI}$ trajectories, and as $\beta$ grows, its area reduces even more until it disappears when $\beta=1$. Finally, only negative $\gamma$ values exist for $\beta > 1$. All these results are consistent with the map of regions in Fig. 3.

\begin{figure}[t]
\includegraphics[width=8cm]{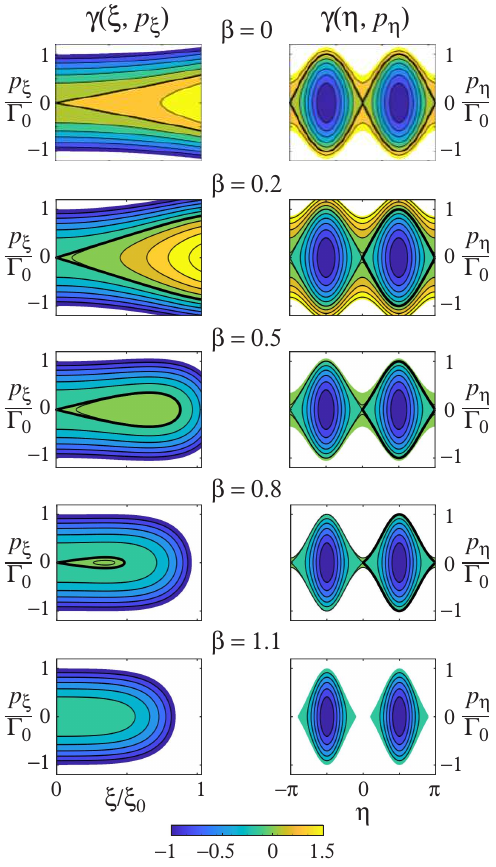}
\caption{\label{Fig4} Poincar\'{e} phase mappings $\left(  \xi,p_{\xi}\right)  $ and $\left(
\eta,p_{\eta}\right)  $ of the billiard with $\gamma_{0}=1.5$ for several
values of the energy parameter $\beta=\{0,0.2,0.5,0.8,1.1\}.$ The iso-$\gamma$
lines are contour lines of the surfaces $\gamma\left(  \xi,p_{\xi}; \beta \right)  $
and $\gamma\left(  \eta,p_{\eta}; \beta \right) $ obtained from Eqs. (\ref{pe}) and
(\ref{pn}), respectively.}
\end{figure}

As $\beta$ increases,
the two lobes of the angular map $\left(  \eta,p_{\eta}\right)  $ become
thinner and thinner until they separate definitively for $\beta>1$. The particle's motion can be traced along a specific iso-$\gamma$ curve in phase
space. For example, a libration motion corresponds to a closed orbit in the
plane $\left(  \eta,p_{\eta}\right)  $ moving in a finite interval of the
coordinate $\eta$ between both hyperbolic caustics.

In the radial map $\left(  \xi,p_{\xi}\right),$ the particle moves towards
the boundary in the upper half-space $p_{\xi}>0;$ conversely, it travels in
the direction of the interfocal line when $p_{\xi}<0.$ The reflections of the
particle at the boundary correspond to changes $+p_{\xi}\rightarrow-p_{\xi}$
that are represented in the map $\left(  \xi,p_{\xi}\right)  $ by a vertical
jump along the line $\xi=\xi_{0}$ connecting the upper and the lower level
curves. The orbits in the map $\left(  \xi,p_{\xi}\right)  $ are always
circulated clockwise.

For arbitrary values of $\gamma$ and $\beta,$ if the particle touches the
billiard boundary, its trajectory is open in general. That is, the particle
never returns to the starting point with the initial momentum. Thus, after
many bounces, it \textit{fills} densely the region bounded by the boundary and
the caustics. For specific values of $\gamma$ and $\beta$, the trajectory can
close and form a self-intersecting polygon (whose sides are elliptic arcs)
inscribed about the caustics and the elliptic wall. In the next section, we
will derive the conditions to get periodic trajectories in the billiard.

\section{Periodic trajectories}

The action variables for the canonical coordinates are \cite{FetterBOOK,GoldsteinBOOK}%
\begin{equation}
J_{\xi}=\frac{1}{2\pi}{\oint}p_{\xi}\left(  \xi\right)  \mathrm{d}\xi,\qquad
J_{\eta}=\frac{1}{2\pi}{\oint}p_{\eta}\left(  \eta\right)  \mathrm{d}\eta,
\end{equation}
where the integrals are carried out over a complete period of the coordinates
$\xi$ and $\eta.$ Replacing $p_{\xi}\left(  \xi\right)  $ and $p_{\eta}\left(
\eta\right)  $ from Eqs. (\ref{pe}) and (\ref{pn}) we get
\begin{align}
J_{\xi}\left(  \gamma,\beta\right)   &  =\frac{c}{2\pi}{\oint}\mathrm{d}%
\xi\sqrt{\sinh^{2}\xi\left(  1-\beta\cosh^{2}\xi\right)  -\gamma},\label{je}\\
J_{\eta}\left(  \gamma,\beta\right)   &  =\frac{c}{2\pi}{\oint}\mathrm{d}%
\eta\sqrt{\sin^{2}\eta\left(  1-\beta\cos^{2}\eta\right)  +\gamma}. \label{jn}%
\end{align}

Given the values of $\gamma,\beta,$ the actions $J_{\xi}$ and $J_{\eta}$ are proportional to the geometric area enclosed by the corresponding orbits on the Poincar\'{e} maps shown in Fig. 4.

According to the types of motion discussed in Sect. \ref{Sect_Traj} and the phase maps in Fig. 4, the closed integrals become open integrals whose limits are
\begin{equation}%
\begin{array}
[c]{ccccccc}%
\begin{array}
[c]{c}%
J_{\xi}\\
\mathcal{R}\text{-type}%
\end{array}
&  &
\begin{array}
[c]{c}%
J_{\xi}\\
\mathcal{L}\text{-type}%
\end{array}
&  &
\begin{array}
[c]{c}%
J_{\eta}\\
\mathcal{R}\text{-type}%
\end{array}
&  &
\begin{array}
[c]{c}%
J_{\eta}\\
\mathcal{L}\text{-type}%
\end{array}
\\%
\displaystyle
{2}\int_{\xi_{C}}^{\xi_{0}}, &  &
\displaystyle
{2}\int_{0}^{\xi_{0}}, &  &
\displaystyle
{4}\int_{0}^{\pi/2}, &  &
\displaystyle
{4}\int_{\eta_{C}}^{\pi/2},
\end{array}
\label{open}%
\end{equation}
where $\xi_{C}$ and $\eta_{C}$ are given by Eqs. (\ref{uCL}) and (\ref{vcL}), respectively.

\subsection{The winding number function}

The winding number $w$ of the system is the ratio of the angle variables
$\omega_{\xi}$ and $\omega_{\eta}$ conjugate to the actions, namely%
\begin{equation}
w=\frac{\omega_{\eta}}{\omega_{\xi}}=\frac{%
\displaystyle
\frac{\partial H}{\partial J_{\eta}}}{%
\displaystyle
\frac{\partial H}{\partial J_{\xi}}}=\frac{\partial J_{\xi}}{\partial J_{\eta
}}=\frac{\left\vert
\displaystyle
\frac{\partial J_{\xi}}{\partial\gamma}\right\vert }{\left\vert
\displaystyle
\frac{\partial J_{\eta}}{\partial\gamma}\right\vert }, \label{condi}%
\end{equation}
with $H$ being the Hamiltonian. Clearly, the winding number is a function of the constants $\left(  \gamma,\beta\right)  .$

The derivatives of the actions with respect to $\gamma$ are
\begin{align}
\frac{\partial J_{\xi}}{\partial\gamma}  &  =-\frac{c}{4\pi}{\oint}%
\frac{\mathrm{d}\xi}{\sqrt{\sinh^{2}\xi\left(  1-\beta\cosh^{2}\xi\right)
-\gamma}},\label{dje}\\
\frac{\partial J_{\eta}}{\partial\gamma}  &  =\frac{c}{4\pi}{\oint}%
\frac{\mathrm{d}\eta}{\sqrt{\sin^{2}\eta\left(  1-\beta\cos^{2}\eta\right)+\gamma}}, \label{djn}%
\end{align}
where the closed integrals are replaced by the corresponding open integral in
Eq. (\ref{open}) depending on the particular case.

Carrying out the changes of variable $u=\sinh^{2}\xi$ and $v=\sin^{2}\eta$,
the integrands of Eqs. (\ref{dje}) and (\ref{djn}) are expressed in terms
of square roots of fourth-order polynomials in the variables $u$ and $v$. This
allows us to write explicit results utilizing the incomplete $F(\phi,k)$ and
the complete $K\hspace{-0.1cm}\left(  k\right)  $ elliptic integrals of the
first kind \cite{GradshteynBOOK,ByrdBOOK}
\begin{equation}
F(\phi,k)=\int_{0}^{\phi}\hspace{-0.2cm}\frac{\mathrm{d}\theta}{\sqrt
{1-k^{2}\sin^{2}\theta}},\quad K\hspace{-0.1cm}\left(  k\right)  =F\left(
\frac{\pi}{2},k\right)  . \label{EI1}%
\end{equation}

We obtain for the rotational ($\mathcal{R}$) and librational ($\mathcal{L}$)
motions
\begin{align}
\frac{\partial J_{\xi}}{\partial\gamma}  &  =%
\begin{cases}%
\displaystyle
-\frac{c}{2\pi}\frac{h}{\sqrt{\mathcal{D}}}F\left(  \phi_{1},h\right)
,\smallskip & \mathcal{R}\text{-type,}\\%
\displaystyle
-\frac{c}{2\pi}\frac{1}{\sqrt{\mathcal{D}}}F\left(  \phi_{2},\frac{1}%
{h}\right)  , & \mathcal{L}\text{-type,}%
\end{cases}
\label{djee}\\
\frac{\partial J_{\eta}}{\partial\gamma}  &  =%
\begin{cases}%
\displaystyle
\frac{c}{\pi}\frac{h}{\sqrt{\mathcal{D}}}K\hspace{-0.1cm}\left(  h\right)
,\smallskip & \mathcal{R}\text{-type,}\\%
\displaystyle
\frac{c}{\pi}\frac{1}{\sqrt{\mathcal{D}}}K\hspace{-0.1cm}\left(  \frac{1}%
{h}\right)  ,\hspace{0.6cm} & \mathcal{L}\text{-type,}%
\end{cases}
\label{djnn}%
\end{align}
where $\mathcal{D}=\sqrt{(\beta-1)^{2}-4\gamma\beta}$ [Eq. (\ref{D})], and%
\begin{align}
h  &  \equiv\sqrt{\frac{2\mathcal{D}}{1-\beta+\mathcal{D}+2\gamma}}%
,\label{h}\\
\sin\phi_{1}  &  =\frac{1}{\sin\phi_{2}}=\sqrt{\frac{1-\beta+\mathcal{D}%
-2\gamma/\gamma_{0}}{2\mathcal{D}}}. \label{sinf}%
\end{align}

Replacing Eqs. (\ref{djee}) and (\ref{djnn}) into Eq. (\ref{condi}), the
winding number function is given by%
\begin{equation}
w\left(  \gamma,\beta\right)  =\left\{
\begin{array}
[c]{lll}%
\displaystyle
\frac{F\left(  \phi_{1},h\right)  }{2K\hspace{-0.1cm}\left(  h\right)},\smallskip &  & \mathcal{R}\text{ region,}\\%
\displaystyle
\frac{F\left(  \phi_{2},1/h\right)  }{2K\hspace{-0.1cm}\left(  1/h\right)},\smallskip &  & \mathcal{L}\text{ region,}\\
1/2, &  & \mathcal{EI}\text{ and }\mathcal{EO}\text{ regions.}%
\end{array}
\right.  \label{wnf}%
\end{equation}

\begin{figure}[t]
\includegraphics[width=8cm]{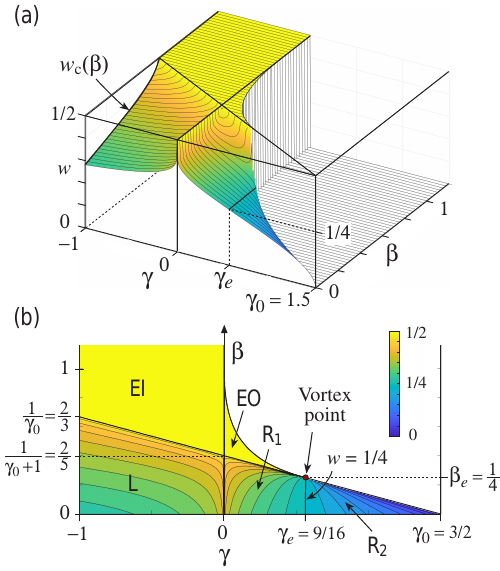}
\caption{\label{Fig5} (a) Winding number surface $w\left(  \gamma,\beta\right)  $ with $\gamma_{0}=1.5$
showing curves iso-$\beta$. The cutoff condition $w_{\mathrm{c}}\left(
\beta\right)  $ is given by Eq. (\ref{wc}). (b) Iso-$w$ level curves of
$w(\gamma,\beta)$\thinspace$=\left\{  0,0.025,0.05,...,0.5\right\}  $. The
winding number cannot be greater than 1/2. (c) Detail of the $\mathcal{R}_{1}$
region. For $\beta>\beta_{e}$, the level curves have two roots of $\gamma,$
which correspond to degenerate orbits in the billiard. For $\gamma=0.4$ we
have $\gamma_{p}=0.1525$ and $\gamma_{q}=0.3284$. }
\end{figure}

The analytical expressions of the winding number function are an important
result of this work. They fully characterize the particle trajectories in the
billiard. Figure 5(a) shows the winding number function $w\left(  \gamma
,\beta\right)  $ for a billiard with $\gamma_{0}=1.5$. Waterfall lines in
subplot 5(a) show the behavior of $w\left(  \gamma,\beta\right)  $ as a
function of $\gamma$ for constant values of $\beta$; that is, they are lines
of constant energy. 

The contour plot in Fig. 5(b) shows the level curves
$w=\left\{  0,0.025,0.05,...,0.5\right\}  $ of the winding number function.
All points $\left(  \gamma,\beta\right)  $ lying on a level curve have the
same winding number. Note the parallelism between the region map in Fig. 3(a)
with the winding number function.

Analysis of the function $w\left(  \gamma,\beta\right)  $ plotted in Fig. 5
reveals the following properties:

\begin{itemize}
\item The winding number reaches the maximum of 1/2 when $\gamma=0$,
corresponding to focal $\mathcal{F}$ trajectories.

\item The elliptical trajectories in regions $\mathcal{EI}$ and $\mathcal{EO}$
have a winding number of 1/2, which tells us that in an elliptic orbit, the
angular coordinate $\eta$ goes through its range once, and the radial
coordinate $\xi$ goes through its range twice.

\item All iso-$w$ lines in the librational region begin at the baseline
$\beta=0$ with negative $\gamma,$ increase monotonically as $\gamma$
decreases, and end at the vertical axis $\gamma=-1.$

\item A librational orbit with energy constant $\beta$ can occur in the
billiard only if its winding number lies in the interval $[w_{\mathrm{c}%
},1/2),$ where
\begin{equation}
w_{\mathrm{c}}=w\left(  -1,\beta\right)  =\frac{1}{\pi}\arcsin\left(
\sqrt{\frac{\gamma_{0}\left(  \beta+1\right)  }{\gamma_{0}+1}}\right)  ,
\label{wc}%
\end{equation}
is the cutoff winding number for librational orbits, see Fig. 5(a).

\item All iso-$w$ lines in the rotational region begin at the baseline
$\beta=0$ with $\gamma>0,$ and converge to the vortex point $\left(
\gamma_{e},\beta_{e}\right)  $ given by Eq. (\ref{gebe}).

\item The winding number at $\gamma=\gamma_{e}$ is constant for all values of
$\beta$, namely
\begin{equation}
w\left(  \gamma_{e},0\leq\beta\leq\beta_{e}\right)  =1/4.
\end{equation}

\item The iso-$w$ line equal to $1/4$ at $\gamma=\gamma_{e}$ divides the
region $\mathcal{R}$ into the subregions $\mathcal{R}_{1}$ and $\mathcal{R}%
_{2}$, as shown in Figs. 3 and 5. All rotational orbits lying in
$\mathcal{R}_{1}$ ($\gamma<\gamma_{e}$) have $w>1/4$, whereas orbits lying in
$\mathcal{R}_{2}$ ($\gamma>\gamma_{e}$) have $w<1/4$ and $\beta<\beta_{e}.$
The level curves in the subregion $\mathcal{R}_{2}$ tend monotonically to the
vortex point, whereas the curves in $\mathcal{R}_{1}$ grow, reach a maximum,
and descend to the vortex.

\item Observe the value of the winding number along the upper border of region $\mathcal{R}$. In zone $\mathcal{R}_{1}$, its value is $w=1/2$, and in zone $\mathcal{R}_{2}$, we have $w=0$. The discontinuity occurs at the vortex point, where the winding number is indeterminate. By adopting the half-maximum convention, the value at the vortex is, by definition, 1/4.
\end{itemize}

\subsection{Degenerate trajectories}

From the behavior of the winding number function shown in Fig. 5, we observe that it is possible to get two different trajectories with the same winding number $w$ and the same $\beta,$ but different values
of $\gamma,$ as long as  $w \in (1/4,1/2)$ and $\beta>\beta_{e}$. To explain this result, in Fig.
6(a), we plot the subregion $\mathcal{R}_{1}$ showing in more detail the
iso-$w$ curves. For example, the horizontal line $\beta=0.3$ is above the
vortex [$\beta_{e}=0.25$] and intersects twice with the contour curve $w=0.4$.
Conversely, the line $\gamma=0.2$ is below $\beta_{e}$; thus, there is only
one cross-point with the curve $w=0.4$. Since all points lying in an
iso-$\beta$ line have the same energy $E$, the trajectories that share the
same $\beta$ and the same winding number $w$ but a different $\gamma$ can be
considered \textit{degenerate trajectories}. By having different $\gamma$, two
degenerate trajectories have different elliptic caustics, periods, and lengths.

To have degenerate trajectories, the point $\left(  \gamma,\beta\right)
$ has to lie within the triangular region defined by the straight lines
$\gamma=0,$ $\beta=\beta_{e},$ and $\beta=\left(  \gamma_{0}-\gamma\right)
/\gamma_{0}(\gamma_{0}+1).$ There is not possible to have degenerate
librational trajectories.


In Fig. 6(b), we show a pair of degenerate trajectories with $\beta=0.3$ and $w=0.4$. 
The location of the corresponding points $\left(  \gamma_{p},\beta\right)  $ and $\left(  \gamma_{q},\beta\right)  $ on the $\left(  \gamma,\beta\right) $ chart is shown in Fig. 6(a). 
The vertices 1 in both trajectories are located at the same point $\eta=75^\circ$ to easily notice that the corresponding vertices between both orbits are located at different positions.
As expected for degenerate trajectories, their caustics, periods, and lengths differ.

\begin{figure}[t]
\includegraphics[width=8cm]{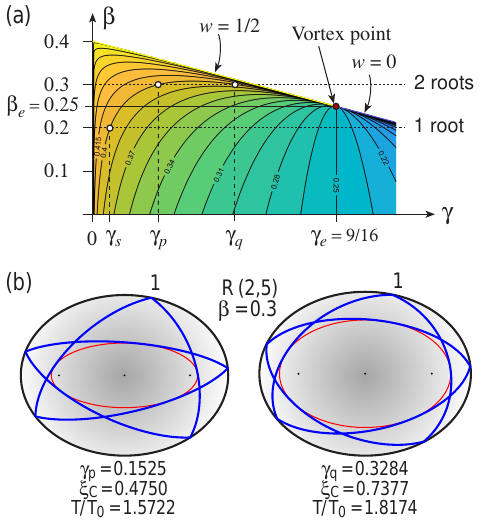}
\caption{Degenerate rotational trajectories $\left(2,5\right)$ for a billiard with $\gamma_0 = 1.5$. Both
orbits have the same winding number $w=0.4,$ the same energy $\beta=0.3,$ but
different constant $\gamma$. The values of the elliptic caustics $\xi_C$ and the periods $T$ in units of $T_0=2\pi/\omega$ are included for each orbit.}
\end{figure}

\subsection{Characteristic equations of periodic trajectories}

The periodic orbits are determined by the condition that the winding number is equal to a rational number%
\begin{equation}
w=w_{m}^{n}=\frac{n}{m}, \label{wnm}%
\end{equation}
where $n$ and $m$ are two integer numbers. The periodic trajectory closes
after $m$ periods of the coordinate $\xi$ and $n$ periods of the coordinate
$\eta$. If $w$ is an irrational number, then the trajectory never closes and
ends up filling the available configuration space inside the billiard.

From Eqs. (\ref{wnf}) and (\ref{wnm}), the characteristic equations to get
periodic orbits $\left(  n,m\right)  $ in the billiard are
\begin{subequations}
\label{CHE1}%
\begin{align}
w_{m}^{n}  &  =\frac{F\left(  \phi_{1},h\right)  }{2K\left(  h\right)  }%
=\frac{n}{m},\qquad\quad\mathcal{R}\text{-type,}\label{Rw}\\
w_{m}^{n}  &  =\frac{F\left(  \phi_{2},1/h\right)  }{2K\left(  1/h\right)
}=\frac{n}{m},\qquad\mathcal{L}\text{-type.} \label{Lw}%
\end{align}
\end{subequations}
These equations have the same structure as the characteristic equations of the elliptic billiard without potential \cite{Waalkens,P05}, but the arguments are different. 

The behavior of the iso-$w$ lines on the plane $\left(  \gamma,\beta\right)  $
is shown in Fig. 5(b). For example, any point $\left(  \gamma,\beta\right)  $
on the iso-$w$ line equal to 0.375 generates a closed path $\left(
n,m\right)  =(3,8)$ that could be rotational or librational. If the value of
$w$ is below the cutoff [Eq. (\ref{wc})], for example $w=0.15$, only
rotational trajectories can exist.

Alternatively, the characteristic equations (\ref{CHE1}) can be inverted by
applying the Jacobian elliptic function $\mathrm{sn}\left(  x,\alpha\right)  $
\cite{ByrdBOOK,NISTBOOK}. For rotational orbits, Eq. (\ref{Rw}) becomes
\begin{subequations}
\label{CHE2}%
\begin{equation}
\mathrm{sn}\left[  \frac{2n}{m}K\left(  h\right)  ,h\right]  =\sqrt
{\frac{1-\beta+\mathcal{D}-2\gamma/\gamma_{0}}{2\mathcal{D}}}. \label{RCE}%
\end{equation}
This equation has real solutions for $m\geq3$ and $n\leq m/2.$ The numbers $m$
and $n$ are the number of bounces at the boundary and the number of turns the
particle makes in a cycle, respectively.

For librational orbits, Eq. (\ref{Lw}) becomes%
\begin{equation}
\mathrm{sn}\left[  \frac{2n}{m}K\left(  \frac{1}{h}\right)
,\frac{1}{h}\right]  =\sqrt{\frac{2\mathcal{D}}{1-\beta+\mathcal{D}%
-2\gamma/\gamma_{0}}}, \label{LCE}%
\end{equation}
\end{subequations}
which has real solutions for $m\geq4$ and $n\leq m/2,$ where $m$ must be an
even integer to have closed librational trajectories.

\begin{figure}[t]
\includegraphics[width=8cm]{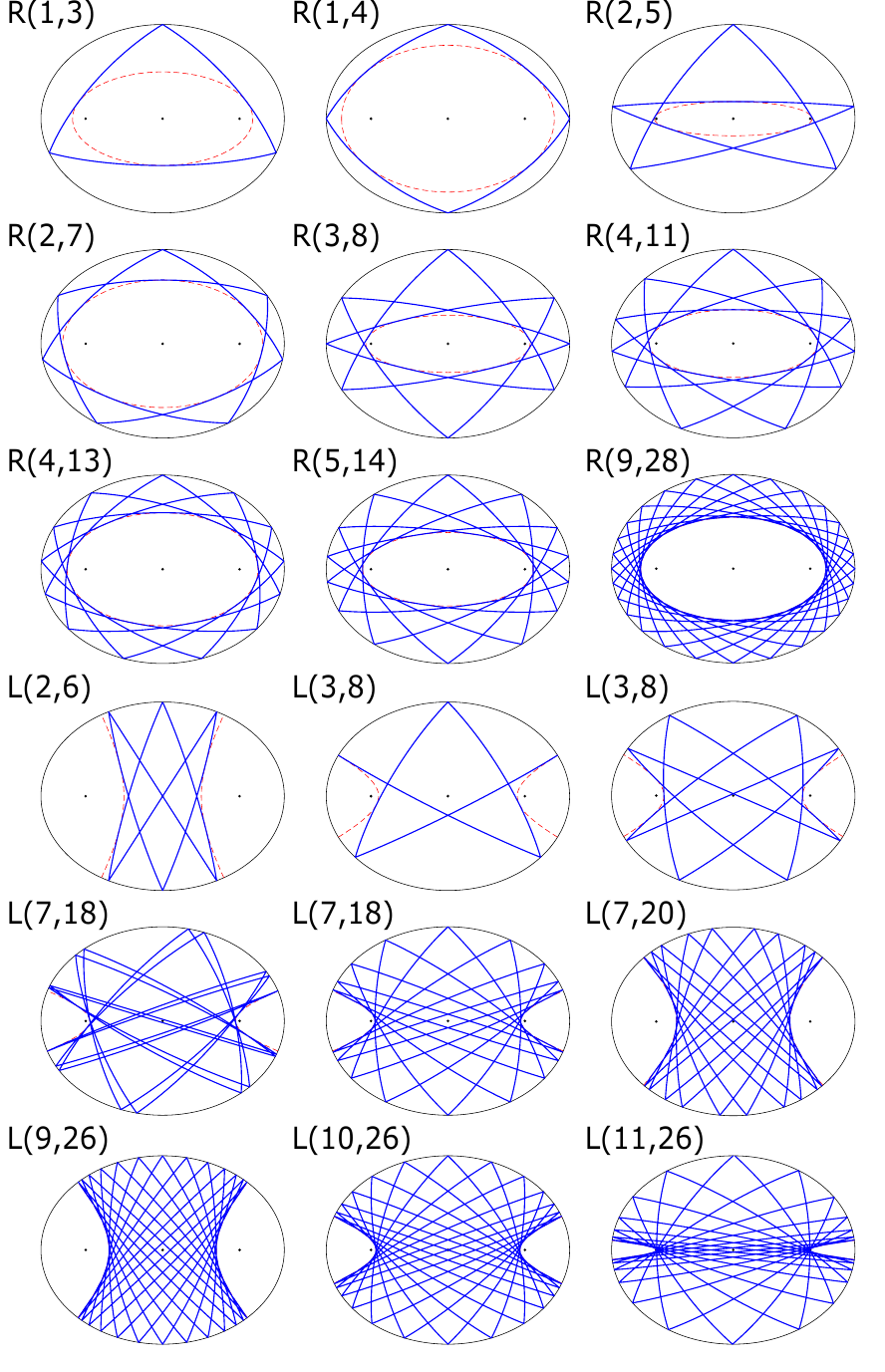}
\caption{\label{Fig6} Rotational ($\mathcal{R}$) and librational ($\mathcal{L}$)
trajectories for a billiard with $\gamma_{m}=1.5.$ Caustics are depicted with
dashed red lines. All orbits have $\beta=0.2.$ }
\end{figure}

The process of determining the periodic trajectories in the billiard is as
follows: For a specific trajectory $\left(  n,m\right)  ,$ either rotational
or librational, locate a point $\left(  \gamma,\beta\right)  $ lying on the
contour line with winding number $w=n/m$. Usually, $\beta$ is proposed (since
it is equivalent to giving the energy), and $\gamma$ is calculated by finding
the root of the corresponding characteristic equation, either using Eq.
(\ref{CHE1}) or (\ref{CHE2}). Determine the value of the caustics evaluating
either Eq. (\ref{uCL}) or (\ref{uCEI}) at the point $\left(  \gamma
,\beta\right)  $. This information lets us know the allowed region where the
particle moves within the billiard. Later, set the coordinates $\left(
\xi,\eta\right)  $ of the starting point of the trajectory; they must be
within the valid region of motion of the particle. Typically, one chooses a
point ($\eta$) on the boundary $\xi=\xi_{0}$. Now, Eqs. (\ref{pV}),
(\ref{pe}), and (\ref{pn}) give the components $p_{\xi}/\sigma$ and $p_{\eta
}/\sigma$ of the initial momentum $\mathbf{p}$, which provides the angle about
the tangent to the boundary of the first segment of the trajectory, namely
$\alpha=\arctan\left(  p_{\xi}/p_{\eta}\right)  $. Calculate the first
elliptic trajectory through the potential and find the impact point with boundary.
Calculate the velocity vector after the bounce considering that the collision
is elastic. From here, it is an iterative process. Trace the complete orbit by
calculating the successive elliptical segments and the collision points on the
boundary. The trajectory will close after $m$ bounces for $\mathcal{R}$-motion
and $2m$ bounces for $\mathcal{L}$-motion.%

Some rotational and librational periodic orbits are depicted in Fig. 7 for a
billiard with $\gamma_{0}=1.5.$ For rotational orbits, $m$ is either the
number of bounces at the boundary or the number of sides, and $n$ is the
number of turns around the interfocal line in a cycle. For librational orbits,
$2m$ is the number of reflections at the boundary, and $n$ is the number of
times the trajectory touches the caustics. In most examples, we select the
upper covertex as the starting point of the trajectory, which produces
symmetric orbits about the y-axis. The topologies of the rotational
trajectories are straightforward, but in the librational case, interesting
phenomena can occur. For example, the orbit $L(3,8)$ is shown twice; in the
first image, the particle bounces perpendicularly on the border, and then it
returns by the same path to complete the trajectory; in the second, we choose
another starting point to unfold the trajectory. The path $L(7,18)$ is also
shown twice to illustrate that symmetric or non-symmetric librational paths
around the y-axis can be obtained. Finally, in the last line, we show three
orbits with $m=26$ but different $n$ to show the effect of the gradual
variation of the winding number on the hyperbolic caustics.

The starting point of a given trajectory $\left(  n,m\right)  $ does not
affect the calculation of the constants of motion $\left(  \gamma
,\beta\right)  $. Thus, we can choose any point of the boundary, within the
allowed region, as the first vertex for constructing the polygonal
trajectory. Moving the initial point along the boundary generates different
orbits with the same number of sides and, as we will see, the same period and
length. Indeed, in an integrable system, the periodic tori are not isolated
but form a continuous family that fills the configuration space.

\subsection{Period of the periodic orbits}

In the simple case when the particle does not hit the boundary and its
trajectory becomes an ellipse (winding number $w=1/2$), it is clear that the
period of the orbit is simply
\begin{equation}
T_{0} = 2\pi/\omega.
\end{equation}
$T_0$ can be considered as the characteristic constant of time of the system.

On the other hand, when the particle describes a polygonal trajectory, the
calculation of the period is much more complex. Fortunately, a general
expression for the period $T_{m}^{n}$ of the trajectory $\left(  n,m\right)  $
can be derived starting from the definition of the Hamiltonian in terms of canonical
coordinates and momenta, i.e., $H=\sum_{j}p_{j}\dot{q}_{j}-L,$ where $L$ is
the Lagrangian, and the over-dot means time derivative. Integrating with
respect to time over one cycle yields%
\begin{subequations}
\begin{align}
\sum_{j}\oint p_{j}q_{j}  &  =\oint H~\mathrm{d}t+\oint L~\mathrm{d}t,\\
2\pi\left(  mJ_{\xi}+nJ_{\eta}\right)   &  =ET_{m}^{n}+S.
\end{align}
\end{subequations}
where $S$ is the action, and $J_{\xi}$ and $J_{\eta}$ are defined by Eqs.
(\ref{je}) and (\ref{jn}). Because $S$ is constant for a specific trajectory,
partial derivation with respect to the energy yields%
\begin{equation}
T_{m}^{n}=2\pi\left(  m\frac{\partial J_{\xi}}{\partial E}+n\frac{\partial
J_{\eta}}{\partial E}\right)  . \label{Tnm}%
\end{equation}
which is the desired expression.

The evaluation of Eq. (\ref{Tnm}) is laborious, but the result can be
expressed in terms of the incomplete $\Pi(\phi,n,k)$ and complete $\Pi(n,k)$
elliptic integrals of the third kind \cite{GradshteynBOOK,ByrdBOOK}
\begin{subequations}
\label{EI3}%
\begin{align}
\Pi(\phi,n,k)  &  =\int_{0}^{\phi}\hspace{-0.2cm}\frac{\mathrm{d}\theta
}{\left(  1-n\sin^{2}\theta\right)  \sqrt{1-k^{2}\sin^{2}\theta}},\\
\Pi(n,k)  &  =\Pi\left(  \pi/2,n,k\right)  .
\end{align}
\end{subequations}

For rotational ($\mathcal{R}$) trajectories, we get%
\begin{multline}
T_{m}^{n}=T_{0}\frac{h}{\pi}\sqrt{\frac{\beta}{\mathcal{D}}}\left\{  mu_{-}%
\Pi\left(  \phi_{1},\frac{1}{s_{+}},h\right)  \right.  +\label{TnmR}\\
\left.  2n\left[  \left(  1-v_{-}\right)  \Pi\left(  \frac{1}{v_{-}},h\right)
+v_{-}K\left(  h\right)  \right]  \right\}  ,
\end{multline}
where $\mathcal{D}$ is given by Eq. (\ref{D}), $u_{\pm}$ by (\ref{umm}),
$v_{\pm}$ by (\ref{vmm}), $h$ by (\ref{h}), $\phi_{1,2}$ by (\ref{sinf}), and
\begin{equation}
s_{\pm}\equiv\frac{\mathcal{D}-\beta\pm1}{2\mathcal{D}}.
\end{equation}

For librational ($\mathcal{L}$) trajectories we get%
\begin{multline}
T_{m}^{n}=T_{0}\frac{1}{\pi}\sqrt{\frac{\beta}{\mathcal{D}}}\times
\label{TnmL}\\
\left\{  mu_{-}\left[  F\left(  \phi_{2},\frac{1}{h}\right)  -\Pi\left(
\phi_{2},s_{+},\frac{1}{h}\right)  \right]  \right.  +\\
\left.  2n\left[  \left(  1-v_{-}\right)  \Pi\left(  s_{-},\frac{1}{h}\right)
+v_{-}K\left(  \frac{1}{h}\right)  \right]  \right\}  .
\end{multline}

Equations (\ref{TnmR}) and (\ref{TnmL}) are formidable; they allow us to
evaluate the period of a periodic orbit in the billiard analytically. We have
compared the results of these equations with those obtained using numerical
simulations of the particle moving in the billiard, and the discrepancy is
less than $10^{-10}$.%

\begin{figure}[t]
\includegraphics[width=8cm]{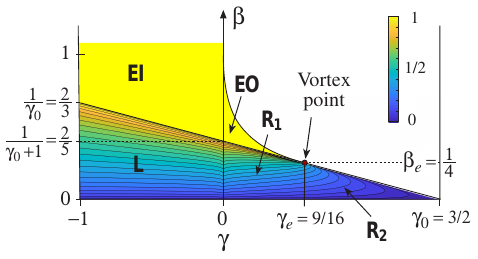}
\caption{Period function $T_{m}^{n}/T_{0}=mf\left(  \gamma,\beta\right)  $
with $m=2.$ The period in the regions $\mathcal{EI}$\ and $\mathcal{EO}$ is
$T_{0},$ as expected.}
\end{figure}

If we now extract the index $m$ from Eq. (\ref{Tnm}) and use the definition of
winding number $w_{m}^{n}=n/m$, the period writes as $T_{m}^{n}=m~2\pi\left(
\partial J_{\xi}/\partial E+w_{m}^{n}\partial J_{\eta}/\partial E\right)  .$
From here, we can infer that the expressions for the period can be written in
the normalized form
\begin{equation}
T_{m}^{n}/T_{0}=m\tau\left(  \gamma,\beta\right)  ,
\end{equation}
where $\tau\left(  \gamma,\beta\right)  $ is a dimensionless function that
only depends on the constants of motion $\left(  \gamma,\beta\right)  $ and is
valid in the regions $\mathcal{L}$ and $\mathcal{R}$.

The behavior of the period function $T_{m}^{n}/T_{0}=m\tau\left(  \gamma
,\beta\right)  $ with $m=2$ is illustrated in Fig. 8. The image shows the
curves of constant period in the rotational and librational regions. Setting
$m=2$ ensures the period at the border with the regions EI and EO is
continuous. The period function for a trajectory $\left(  n,m\right)  $ is the
same, except for a scale factor of $m/2.$

Figure 8 reveals other interesting results of the billiard. As it happened in the winding number function in Fig. 5, it is possible to get
two different orbits with the same energy $\beta$ but different $\gamma$ as
long as the points $\left(  \gamma,\beta\right)  $ lie in the upper triangular
zone of the region $\mathcal{R}_{1}$ above $\beta_{e}$. Further analysis of the
contour lines in Fig. 8 reveal that it is also possible to get two different
orbits in the region $\mathcal{R}_{2}$ with the same $\gamma$ and different
energy $\beta$ that share the same period.

Finally, it is worth mentioning that the period of the orbit $(n,m)$ can also
be calculated with
\begin{equation}
T_{m}^{n}=\pi\frac{mJ_{\xi}\left(  \gamma,\beta\right)  +nJ_{\eta}\left(
\gamma,\beta\right)  }{\left\langle E_{k}\right\rangle }%
\end{equation}
where $\left\langle E_{k}\right\rangle $ is the average value of the kinetic
energy in a complete cycle. The result is fully equivalent
to Eqs. (\ref{TnmR}) and (\ref{TnmL}).

\section{Geometric constructions}

As illustrated in Fig. 9, the self-intersecting points of a trajectory
$\left(  n,m\right)  $ with constants of motion $\left(  \gamma,\beta\right)
$ lie on ellipses (and hyperbolae) that are confocal to the boundary. Any
selection of these confocal ellipses can define a new internal billiard that
supports a new sub-trajectory with the same $\left(  \gamma,\beta\right)  $
but different indices $\left(  n^{\prime},m\right)  .$ In the same way, if we
extend the elliptical segments beyond the boundary (as if it did not exist),
we can see that the outer elliptic paths also intersect at confocal ellipses
that could be considered the wall of larger elliptic billiards. This result
applies to both rotational and librational orbits. In the following, we adopt
the term \textit{SI}-ellipses to refer to the confocal ellipses outlined by
the cross-points where the elliptic paths intersect. Let us analyze the
rotational and librational cases separately.

\begin{figure}[t]
\includegraphics[width=6cm]{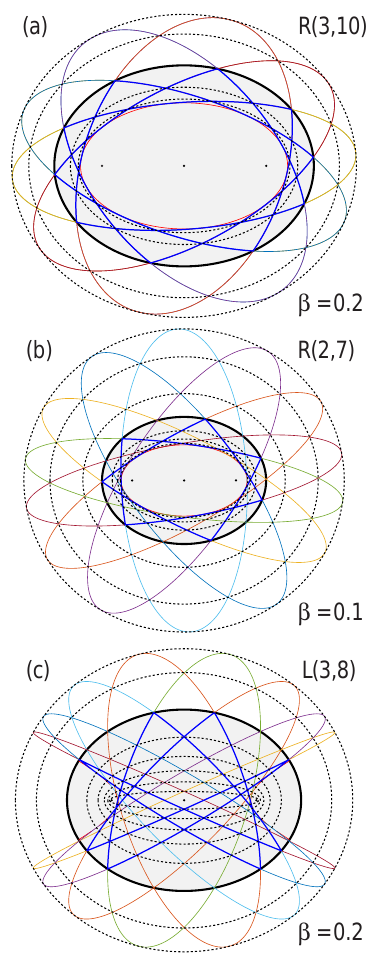}
\caption{Self-Intersecting ellipses (dashed lines) for (a) rotational
trajectory (3,10) with $\beta=0.2$ and $\gamma=0.403$; (b) rotational
trajectory (2,7) with $\beta=0.1$ and $\gamma=0.402,$ and (c) librational
trajectory (3,8) with $\beta=0.2$ and $\gamma=-0.158.$}
\end{figure}

\subsection{Rotational orbits}

For $\mathcal{R}$ orbits $\left(  n,m\right)  $ with even $m,$ there are $m/2$
\textit{SI}-ellipses distributed as follows: $\left(  n-1\right)  $ internal
ellipses, $1$ corresponding to the boundary, and $\left(  m/2-n\right)  $
external ellipses. These results are exemplified in Fig. 9(a) for a rotational
orbit with $\left(  n,m\right)  =\left(  3,10\right)  $ and $\beta=0.2.$ The
winding numbers $w^{\prime}=n^{\prime}/m$ of the trajectories $\left(
n^{\prime},m\right)  $ formed by the family of \textit{SI-}ellipses are
\begin{equation}
w^{\prime}=\overset{\left(  n-1\right)  \text{ internal}}{\overbrace{\frac
{1}{m},\cdots,\frac{n-1}{m}}},\overset{\text{1 boundary}}{\overbrace{\frac
{n}{m}}},\overset{\left(  m-2n\right)  /2\text{ external}}{\overbrace
{\frac{n+1}{m},\cdots,\frac{m/2}{m}}},\label{np}%
\end{equation}
where the index $n^{\prime}$ is the number of turns around the interfocal line
that the particle makes in a complete orbit with $m$ bounces. Note that the
amount of \textit{SI}-ellipses is defined exclusively by the number $m$ of
bounces at the boundary. Thus, as long as $m$ is constant, we can gradually
vary the physical parameters of the problem and the number of \textit{SI}%
-ellipses does not change.

The elliptical radii $\xi_{0}$ of the \textit{SI}-ellipses can be found with
the characteristic equation (\ref{RCE}). So far, we have considered the
winding number $w$ as a function of $\left(  \gamma,\beta\right)  $, and the
goal has been to calculate a pair $\left(  \gamma,\beta\right)  $ for a given
$w=n/m.$ Now, the problem can be inverted and formulated as follows: Given
$\left(  \gamma,\beta\right)  $ and the winding numbers $w^{\prime}=n^{\prime
}/m$, find the values of $\gamma_{0}$ that satisfy Eq. (\ref{RCE}).

Solving $\gamma_{0}$ from Eq. (\ref{RCE}), we get
\begin{align}
\gamma_{0}^{\left(  n^{\prime},m\right)  }  &  =\sinh^{2}\xi_{0}^{\left(
n^{\prime},m\right)  }\nonumber\\
&  =\frac{2\gamma}{1-\beta+\mathcal{D}-2\mathcal{D~}\mathrm{sn}^{2}%
\mathcal{\hspace{-0.08cm}}\left[  2w^{\prime}K\left(  h\right)  ,h\right]  },
\label{y0R}%
\end{align}
where $w^{\prime}=n^{\prime}/m$ takes the values according to Eq. (\ref{np}).
Note that $\mathcal{D}$ [Eq. (\ref{D})] and $h$ [Eq. (\ref{h})] depend
exclusively on $\left(  \gamma,\beta\right)  $.

The last \textit{SI}-ellipse with $w=n^{\prime}/m=1/2$, corresponds to the
outermost elliptic caustic formed by the return points of the external
trajectories, see Fig. 9(a). Replacing $n^{\prime}/m=1/2$ into Eq. (\ref{y0R})
and using $\mathrm{sn}\left[  K\left(  h\right)  ,h\right]  =1$, we get the
radius $\xi_{0}^{\left(  m/2\right)  }$ of the extreme caustic
\begin{equation}
\xi_{0}^{\left(  m/2,m\right)  }=\mathrm{arcsinh}\left(  \sqrt{\frac{2\gamma
}{1-\beta-\mathcal{D}}}\right)  .\label{xhi0_om}%
\end{equation}

The construction of the \textit{SI}-ellipses for the case when $m$ is odd is
shown in Fig. 9(b) for $(n,m)=(2,7)$ with $\beta=0.1$. Note that the
cross-points which outline the \textit{SI}-ellipses correspond to the
intersections of $m$ ellipses oriented at different angles. Consequently, the
number of \textit{SI}-ellipses is doubled with respect to the case when $m$ is
even. The winding numbers $w^{\prime}=n^{\prime}/m$ of the corresponding
orbits are%
\begin{equation}
w^{\prime}=\overset{\left(  2n-1\right)  \text{ internal}}{\overbrace{\frac
{1}{2m},\cdots,\frac{2n-1}{2m}}},\overset{\text{1 boundary}}{\overbrace
{\frac{2n}{2m}}},\overset{\left(  m-2n\right)  \text{ external}}%
{\overbrace{\frac{2n+1}{2m},\cdots,\frac{m}{2m}}}.\label{nq}%
\end{equation}
The elliptic radii $\xi_{0}^{\left(  n^{\prime},m\right)  }$of the
\textit{SI}-ellipses with odd $m$ can be also determined with Eq. (\ref{y0R})
taking the winding numbers from Eq. (\ref{nq}).

\subsection{Librational orbits}

Figure 8(c) shows the \textit{SI}-ellipses for a librational orbit with
$(n,m)=(3,8)$ and $\beta=0.2.$ In order to have closed trajectories, $m$ is
always an even number for $\mathcal{L}$ trajectories.

The winding numbers $w^{\prime}=n^{\prime}/m$ of the \textit{SI}-ellipses are
given by Eq. (\ref{nq}), but now their elliptical radii $\xi_{0}$ are
calculated by solving the characteristic equation (\ref{LCE}) for $\gamma
_{0,}$ we have%
\begin{equation}
\sinh^{2}\xi_{0}^{\left(  n^{\prime},m\right)  }=\frac{2\gamma}{1-\beta
+\mathcal{D}-2\mathcal{D/}\mathrm{sn}^{2}\left[  2w^{\prime}K\left(
1/h\right)  \right]  }.
\end{equation}
The outermost SI-ellipse of the librational case can be calculated with Eq.
(\ref{xhi0_om}) as well.

Finally, in characterizing the \textit{SI}-ellipses generated by the self-intersections
of the trajectories in the billiard, we have partially solved the
\textit{outer problem}. In this problem, the particle moves outside the
elliptic wall and is attracted toward the origin by the parabolic potential.
The trajectory is created with the particle bouncing off the boundary from the
outside. The goal is to determine the conditions to get rotational or
librational periodic trajectories. The rotational orbits in the outer problem have similar characteristics to the
rotational orbits we reviewed above. Nevertheless, librational trajectories
are somewhat different since the particle cannot go through the wall, so it
can only move above or below the billiard, as shown in Fig. 9. In any case,
the main properties and the basic equations of the outer problem can be inferred
from the inner problem we discussed in this paper.

\section{Conclusions}

In this paper, we characterize the particle trajectories in an elliptic
billiard with an attractive harmonic oscillator potential, emphasizing the analysis of the periodic trajectories.

It was found that there are four main motion scenarios: rotational,
librational, inner elliptical, and outer elliptical. Additionally, there are some 
particular cases, such as rectilinear and focal motions. Two independent constants of motion characterize the particle dynamics: $\beta$, 
associated with the total energy, and $\gamma$, associated with the
angular momenta about the foci and the position $y$ within the billiard. The
different scenarios can be mapped in the $\left(  \gamma,\beta\right)$ plane, which helps to understand the constraints and ranges of the constants
of motion for a particular trajectory to occur.

We derived closed analytical expressions for the winding number function
$w\left(  \gamma,\beta\right)  $ and the characteristic equations to get
periodic trajectories with angular $n$ and radial $m$ indices. These are
expressed in terms of elliptic integrals of the first kind. We found that it
is possible to have two degenerate $(n,m)$ rotational trajectories that share
the same energy $\beta$ but different $\gamma$ values. It is not possible to
get degenerate librational trajectories.

A notable result was the closed expressions of the time period $T_{m}^{n}$ of
rotational and librational orbits. These expressions are written in terms of
elliptic integrals of the third kind. It was also shown that it is possible to obtain two
different rotational trajectories with the same period and $\gamma$ but
different $\beta$ energy.

We analyzed the caustics and ellipses outlined by the self-intersections of an
orbit $\left(  n,m\right)  $ in the billiard, both for intersections occurring inside and outside the elliptical wall.






\providecommand{\noopsort}[1]{}\providecommand{\singleletter}[1]{#1}%
%

\end{document}